\documentclass[sigconf,nonacm,urlbreakonhyphens=false]{acmart}

\usepackage[outline]{contour}
\acmConference[ICSE 2022]{The 44th International Conference on Software Engineering}{May 21–29, 2022}{Pittsburgh, PA, USA}
\usepackage[utf8]{inputenc}

\setcopyright{none}
\renewcommand\footnotetextcopyrightpermission[1]{} 

\usepackage{listings}
\usepackage{subfigure}
\usepackage{xspace}
\usepackage[utf8]{inputenc}
\usepackage[T1]{fontenc}
\usepackage{xcolor}
\usepackage[normalem]{ulem}
\usepackage{footnote}
\usepackage{tablefootnote}
\usepackage{pdfpages}
\usepackage{multicol}
\usepackage{multirow}
\usepackage{enumitem}
\usepackage{highlighter}
\usepackage{tikz}
\usetikzlibrary{calc}
\usepackage[shortcuts]{extdash}

\newcommand{\jigsaw}{\ensuremath{\mathsf{Jigsaw\xspace}}}
\newcommand{\gpt}{\ensuremath{\mathsf{GPT\text{-}3}}}
\newcommand{\lm}{\ensuremath{\mathsf{PTLM\xspace}}}
\newcommand{\lms}{\ensuremath{\mathsf{PTLM}}}
\newcommand{\codex}{\ensuremath{\mathsf{Codex\xspace}}}

\newcommand{\mathC}{\mathbf{C}}

\newcommand{\nameTransform}{Variable Name}
\newcommand{\nameTransforms}{Variable Name transformations}
\newcommand{\argTransforms}{Argument transformations}
\newcommand{\astTransforms}{AST-to-AST transformations}

\newcommand{\semrep}{Semantic Repair}

\newcommand{\ap}{\ensuremath{\mathsf{AutoPandas\xspace}}}
\newcommand{\apsmall}{\ensuremath{\mathsf{AP\xspace}}}

\newcommand{\prose}{\ensuremath{\mathsf{Prose\xspace}}}

\newcommand{\webapp}{\ensuremath{\mathsf{PandasEval1\xspace}}}
\newcommand{\hackathon}{\ensuremath{\mathsf{PandasEval2\xspace}}}
\newcommand{\hackathonOffline}{\ensuremath{\mathsf{PandasEval2\xspace}}}

\newcommand{\hackathonOnlineOne}{\ensuremath{\mathsf{PandasEval2\_S1\xspace}}}
\newcommand{\hackathonOnlineTwo}{\ensuremath{\mathsf{PandasEval2\_S2\xspace}}}

\newcommand{\no}{\ensuremath{\mathsf{NO\text{-}CONTEXT}}}
\newcommand{\random}{\ensuremath{\mathsf{RAND}}}
\newcommand{\tfidf}{\ensuremath{\mathsf{TFIDF}}}
\newcommand{\transformer}{\ensuremath{\mathsf{TRANSFORMER}}}

\newcommand{\algocode}{\ensuremath{\mathsf{CODE}}}
\newcommand{\algobank}{\ensuremath{\mathsf{BANK}}}
\newcommand{\algoedit}{\ensuremath{\mathsf{EDIT}}}

\newcommand{\csone}{\ensuremath{\mathsf{CS1}}}
\newcommand{\cstwo}{\ensuremath{\mathsf{CS2}}}
\newcommand{\tsone}{\ensuremath{\mathsf{TS1}}}
\newcommand{\tstwo}{\ensuremath{\mathsf{TS2}}}

\newcounter{inlineenum}
\renewcommand{\theinlineenum}{\alph{inlineenum}}
\newenvironment{inlineenum}
  {\unskip\ignorespaces\setcounter{inlineenum}{0}%
   \renewcommand{\item}{\refstepcounter{inlineenum}{\textit{\theinlineenum})~}}}
  {\ignorespacesafterend}

\usepackage{algorithm}
\usepackage[noend]{algpseudocode}
\usepackage{listings}
\usepackage{varwidth}
\usepackage{hhline}
\lstalias[]{csharp}[Sharp]{C}
\lstdefinelanguage{dsl}{
morekeywords = {language,feature,using,semantics,learners,int,string,@input,@start,values,let,in,@ref,@values,Tuple,@extern,@output,namespace,bool,std,@id,grammar},
otherkeywords = {:=,=>,:,[]},
sensitive = true,
morecomment = [l]{//},
morestring = [b]',
}

\lstnewenvironment{myverbatim}[1][]{%
  \lstset{
    basicstyle=\ttfamily,
    columns=fullflexible,
    keepspaces=true,
    frame=tb,
    #1
  }%
}{}

\newcommand{\naman}[1]{{\color{purple}#1}}
\newcommand{\spsays}[1]{{\color{red} \textbf{SP:} #1 }}
\newcommand{\skanda}[1]{{\color{orange} #1}}

\newcommand{\out}[1]{{}}

\colorlet{punct}{red!60!black}
\definecolor{background}{HTML}{EEEEEE}
\definecolor{delim}{RGB}{20,105,176}
\colorlet{numb}{magenta!60!black}

\lstdefinelanguage{json}{
    basicstyle=\normalfont\ttfamily,
    numbers=left,
    numberstyle=\scriptsize,
    stepnumber=1,
    numbersep=8pt,
    showstringspaces=false,
    breaklines=true,
    frame=lines,
    backgroundcolor=\color{background},
    literate=
     *{0}{{{\color{numb}0}}}{1}
      {1}{{{\color{numb}1}}}{1}
      {2}{{{\color{numb}2}}}{1}
      {3}{{{\color{numb}3}}}{1}
      {4}{{{\color{numb}4}}}{1}
      {5}{{{\color{numb}5}}}{1}
      {6}{{{\color{numb}6}}}{1}
      {7}{{{\color{numb}7}}}{1}
      {8}{{{\color{numb}8}}}{1}
      {9}{{{\color{numb}9}}}{1}
      {:}{{{\color{punct}{:}}}}{1}
      {,}{{{\color{punct}{,}}}}{1}
      {\{}{{{\color{delim}{\{}}}}{1}
      {\}}{{{\color{delim}{\}}}}}{1}
      {[}{{{\color{delim}{[}}}}{1}
      {]}{{{\color{delim}{]}}}}{1},
}

\definecolor{codegreen}{rgb}{0,0.6,0}
\definecolor{codegray}{rgb}{0.5,0.5,0.5}
\definecolor{codepurple}{rgb}{0.58,0,0.82}
\definecolor{backcolour}{rgb}{0.95,0.95,0.95}

\lstdefinestyle{mystyle}{
    backgroundcolor=\color{backcolour},   
    commentstyle=\color{codegreen},
    keywordstyle=\color{magenta},
    numberstyle=\tiny\color{codegray},
    stringstyle=\color{codepurple},
    basicstyle=\ttfamily\footnotesize,
    breakatwhitespace=false,         
    breaklines=true,                 
    captionpos=b,                    
    keepspaces=true,                 
    numbers=left,                    
    numbersep=5pt,                  
    showspaces=false,                
    showstringspaces=false,
    showtabs=false,                  
    tabsize=2,
    numbers=none,
    language=Python,
    frameshape={RYR}{Y}{Y}{RYR},
    xleftmargin=.015\textwidth, 
    xrightmargin=.015\textwidth
}

\begin{document}

\title{Jigsaw: Large Language Models meet Program Synthesis}

\author{Naman Jain}
\email{t-namanjain@microsoft.com}
\affiliation{\institution{Microsoft Research}\city{Bangalore}\country{India}}

\author{Skanda Vaidyanath}
\authornote{Work done by author during internship at Microsoft Research, India}
\email{svaidyan@stanford.edu}
\affiliation{\institution{Stanford University}\city{Stanford}\country{USA}}

\author{Arun Iyer}
\email{ariy@microsoft.com}
\affiliation{\institution{Microsoft Research}\city{Bangalore}\country{India}}

\author{Nagarajan Natarajan}
\email{nagarajn@microsoft.com}
\affiliation{\institution{Microsoft Research}\city{Bangalore}\country{India}}

\author{Suresh Parthasarathy}
\email{supartha@microsoft.com}
\affiliation{\institution{Microsoft Research}\city{Bangalore}\country{India}}

\author{Sriram Rajamani}
\email{sriram@microsoft.com}
\affiliation{\institution{Microsoft Research}\city{Bangalore}\country{India}}

\author{Rahul Sharma}
\email{rahsha@microsoft.com}
\affiliation{\institution{Microsoft Research}\city{Bangalore}\country{India}}

\begin{abstract}
    Large pre-trained language models such as GPT-3~\cite{GPT3}, Codex~\cite{Codex}, and Google's language model~\cite{GoogleMNN} are now capable of  generating code from natural language specifications of programmer intent. We view these developments with a mixture of optimism and caution. On the optimistic side, such large language models have the potential to improve productivity by providing an automated AI pair programmer for every programmer in the world. On the cautionary side, since these large language models do not understand program semantics, they offer no guarantees about quality of the suggested code. In this paper, we present an approach to augment these large language models with post-processing steps based on program analysis and synthesis techniques, that understand the syntax and semantics of programs. Further, we show that such techniques can make use of user feedback and improve with usage. We present our experiences from building and evaluating such a tool \jigsaw, targeted at synthesizing code for using Python Pandas API using multi-modal inputs.
    Our experience suggests that as these large language models
    evolve for synthesizing code from intent, \jigsaw~ has an important role to play in improving the accuracy of the systems.
    
    \out{accuracy and quality of the code produced.}
    
    \out{
        Meeting minutes :-
        \begin{itemize}
            \item Explain motivation for pandas
            \item Pandas is hard program synthesis domain - DSL is very big (it is basically python grammar)
            \item in case we want to distinguish ourselves from AP -- we primarily use text and our domain is larger than just dataframe functions - we cover series / python constructs likes lists, dicts, lambda functions
            \item Explain multi-modal specification and motivate use based on security and performance of system
            \item Motivate lifelong learning - potentially with an example? Tie in with large grammar and handcrafting it is not possible 
            \item At some place we would also want to emphasize that lifelong learning is difficult since we will deal with noisy data. For instance users might have written a bad query for an example or  submitted a bad solution; In the worst case we might also get adversarial users. In this work we assume that to not happen? 
            \item Should we mention best case scenario of migrating Jigsaw to a new domain -- Initialize a few prompts and then let generators/refazer/transformation learn over time? (i.e. not a lot of manual intervention)
            \item \textbf{GPT-3, CODEX, MNN will keep on arriving for text-to-code synthesis ; these models are not perfect and blatantly make errors. Therefore one can either sit on sidelines, train new model, fine-tune-model or do post-processing on top of it }. We show that latter is a possibility and gives performance improvements. \textbf{To be rephrased nicely : This work is not about how we built the best program synthesis system but giving a notion of how one can go about building so with only black-box-access to a foundation model / \lm}
            \item Emphasize that \lm do not understand the code but rather just do some retrieval and some generalization. (based on google-paper's results on using transformer to execute code)
            \item Post-processing units are independent improvable search based components; They are guaranteed (if implemented correctly) to make necessary changes unlike gradient descent based learning ; So our way of improving gpt3 is sort of explainable and a good idea for security fixes? Example : We have one unit for arg seelction, one for syntax fixes, one for variable name fixes etc. and these can be improved independently without interfering with each other. this is much harder to achieve with end to end learning where improving one aspect may interfere with the other. 
            \item At the end of the day, these large models dont have understanding of code and is basically just memorizing so there is no guarantee of performance on a wide range of domains. There is only empirical guarantee. We can help take this a step further with some explainable post processing
            \item TODO in discussion section -- "JigSaw in long term when language models's also improve significantly" ; Open ended question -- can we learn even more general transformations  
        \end{itemize}
    }
\end{abstract}

\maketitle

\section{Introduction}
Pre-trained large language models (\lm) such as GPT-3~\cite{GPT3} are finding pervasive applications in Natural Language Processing (NLP), as a general purpose platform to solve many NLP tasks. Recent efforts show that \lm s can generate code from natural language prompts, by associating documentation text with code from a large training set~\cite{Codex, GoogleMNN, Copilot}. This presents a new avenue for program synthesis. However, \lm s  do not ``understand'' either the syntax or semantics of the code, and treat code as text~\cite{GoogleMNN}. Consequently, the code produced by such models has no guarantees of correctness or quality. Hence, any system that uses such \lm s to generate code will need to augment it with program analysis and program synthesis modules to ensure correctness. 
In this paper, we present the design and empirical evaluation of such a multi-modal program synthesis system called \textbf{Jigsaw}, which is targeted specifically at synthesizing code for using large and complex APIs.

\begin{figure}
\fbox{\includegraphics[width=0.8\linewidth]{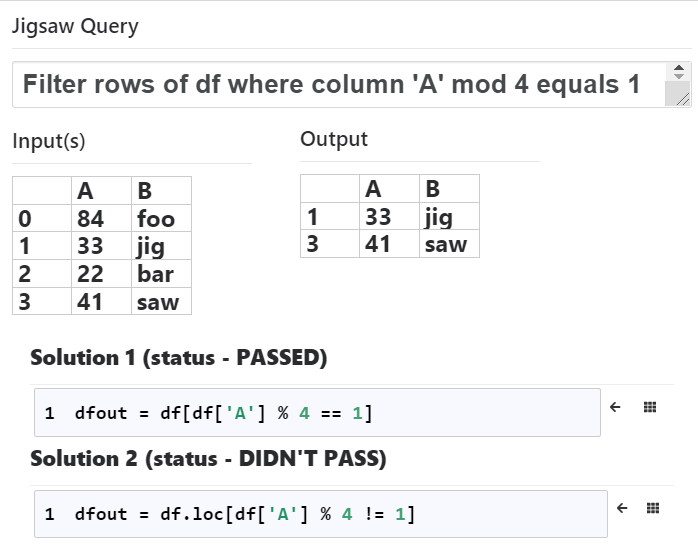}}
\vspace{-3ex}
\caption{Multi-modal problem specification in~\jigsaw}
\label{fig:problem-instance}
\end{figure}

\begin{figure*}[t]
\begin{tikzpicture}[every node/.style={inner sep=5}]
\node[draw, rectangle, ultra thick] (n0) {\begin{tabular}{c} Pre-process \\ inputs \end{tabular}} ;
\node[left of=n0, xshift=-2.8cm, yshift=1cm] (i1) {\begin{tabular}{l} Natural language\end{tabular}} ;
\node[left of=n0, xshift=-2.8cm] (i2) {\begin{tabular}{l} Input/output \\ examples \end{tabular}} ;
\node[left of=n0, xshift=-2.8cm, yshift=-1cm] (i3) {\begin{tabular}{l} Other specifications \\ e.g. Assertions \end{tabular}} ;
\node[left of=n0, xshift=-2.8cm, yshift=-2cm] (i4) {\includegraphics[width=.06\textwidth]{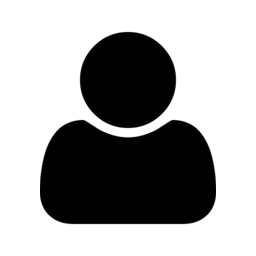}} ;
\node[draw, rectangle, ultra thick, right of=n0, xshift=2.5cm] (n1) { \begin{tabular}{c} Pre-trained \\ language model \lm \end{tabular} } ;
\node[draw, rectangle, ultra thick, right of=n1, xshift=2.5cm] (n2) {\begin{tabular}{c} Post-process \\ outputs \end{tabular}} ;
\node[right of=n2, xshift=2.5cm] (n3) { \begin{tabular}{c} Correct program  \\ (edited by users) \end{tabular} } ;
\node[above of=n3] (i4) {\includegraphics[width=.06\textwidth]{images/user-icon.png}} ;
\node[draw, rectangle, ultra thick, below of=n3, yshift=-0.5 cm] (n4) {Learning from user feedback} ;

\path[draw,->] (i1) -- (n0) ;
\path[draw,->] (i2) -- (n0) ;
\path[draw,->] (i3) -- (n0) ;
\path[draw,->] (n0) -- (n1) ;
\path[draw,->] (n1) -- (n2) ;
\path[draw,->] (n2) -- (n3) ;
\path[draw,->] (n3) -- (n4) ;
\path[draw,->] (n4) -| ($ (n0.south) + (0cm, 0cm) $) ;
\path[draw,->] (n4) -| ($ (n2.south) + (0cm, 0cm) $) ;
\end{tikzpicture}
 \vspace{-5ex}
\caption{Architecture of Jigsaw}
\label{fig:jigsaw-architecture}
\vspace{-0.3cm}
\end{figure*}

\jigsaw\ is multi-modal (as depicted in Figure~\ref{fig:problem-instance}) in the sense that it can ingest input as (1) a natural language string expressing intent and (2) a set of test cases, or input-output examples, and produces a code snippet as output. Future incarnations may be designed to accept other modes of input as well.
The architecture of \jigsaw\ is shown in Figure~\ref{fig:jigsaw-architecture}. The pre-processing module converts the natural language intent into a customized query to send to the \lm. The post-processing module performs syntactic and semantic checks, and performs transformations on the code produced by the \lm, ensuring that the code passes the supplied test cases and other quality checks. The transformations are specifically designed to correct common and recurring errors made by \lm s, such as referencing errors (where the code references variable names incorrectly), argument errors (where the code invokes the correct API, but with incorrect arguments), and a class of semantic errors (which can be corrected by learning \astTransforms). Section~\ref{sec:overview}~shows concrete examples of such errors, and Section~\ref{sec:architecture}~shows how the transformations correct such errors.
\jigsaw\ learns from usage by incorporating user feedback into both pre-processing and post-processing modules, and learns from user engagements to improve its overall quality. Our experiments show how Jigsaw is able to learn from past usage to improve future performance.

The current version of \jigsaw\ is designed and evaluated to synthesize code for the Python Pandas API~\cite{PandasAPI}.  However, the principles behind the design of \jigsaw\ are general, and the design can be extended to other libraries and programming languages as well. We create a user interface for \jigsaw\ using a Jupyter notebook~\cite{jupyter} extension. The extension can be invoked using  a magic command, and invocation of the command creates a sidebar window with a \jigsaw\ card for each invocation. The \jigsaw\ card allows users to supply and edit inputs to the system, inspect the results and copy the desired output back into the main notebook window.

We evaluate \jigsaw\ in terms of the overall accuracy, as well as accuracy of the components of the pre-processing and post-processing modules, on two datasets we created: \webapp, created by the authors of this paper, and \hackathon, created by $25$ users during a hackathon, where participants were given points for solving Python Pandas tasks using \jigsaw. The hackathon was conducted across two sessions (details in Section~\ref{sec:benchmarks}). We used user feedback from the first session to improve the pre-processing and post-processing modules of \jigsaw, and found users were about to solve about $10$\% more tasks in the second session, due to learning improvements from the first session.

We instantiate~\jigsaw\ with two state-of-the-art~\lm{s}:~\gpt~\cite{GPT3} and \codex~\cite{Codex}, and present comprehensive evaluations in Section~\ref{sec:experiments}. We show the overall improved performance of \jigsaw~compared to baselines and state-of-the-art code synthesis frameworks on the two datasets, as well as gains due to learning from user feedback over time.

In summary, this paper makes the following contributions:
\begin{itemize}
    \item
    We present an architecture to perform code synthesis by augmenting black-box \lm{}s with program analysis and synthesis-based techniques and multi-modal specifications.
    We have implemented the architecture in a tool called \jigsaw. We have developed a Jupyter notebook extension that allows users to interact with the system seamlessly.
    
    \item
    We characterize common classes of errors made by \lm{}s, namely, reference errors, argument errors, and semantic errors. Motivated by these errors, we have designed program analysis and synthesis techniques in \jigsaw\ to fix such errors in code snippets produced by \lm{}s. We have also designed techniques to learn from user feedback and improve with usage.
    
    \item 
    We have created two Pandas datasets with multi-modal specifications (provided in the supplementary material, and to be released for community use).  Using two state-of-the-art \lm{}s, we show that~\jigsaw\ yields significantly higher accuracy compared to baselines on the two datasets. 
    
\end{itemize}

Our hypothesis is that even as \lm{}s for code improve, systems such as \jigsaw\ that perform pre-processing and post-processing modules will be crucial to improve user experience, and enhance the quality of the output produced. This is because \lm{}s inherently do not understand the syntax or semantics of code they generate, so we expect gaps to remain between \lm~output and user expectation. Tools based on program analysis and synthesis techniques that understand the code and API syntax and semantics can address these gaps better than generic \lm{}s. We discuss how to design pre-processing and post-processing modules in a general manner, so that \jigsaw\ can work for any language and any API. 

\section{Jigsaw Overview}
\label{sec:overview}
\jigsaw~is a \textit{multi-modal}, \textit{interactive} code synthesis system where (a) the user specifies intent via a combination of natural language description and test cases (i.e., input-output (I/O) examples); and (b) the user interacts with the system via a friendly and seamless interface integrated with the programming environment. The interactive aspect of \jigsaw~is crucial for the developer to refine the possibly ambiguous intent specification as well as for the system to gather useful feedback for improving the components. In this section, we highlight some of the challenges in using general-purpose \lms{}s for specific domains with example queries, which directly motivates the design of our \jigsaw~code synthesis pipeline.

\subsection{~\jigsaw~design principles} We treat large language models as black-box, i.e., we can only query them. This is a reasonable assumption since the premise that the expertise and means to fine-tune the models is out of reach for most users. This design choice is motivated by three reasons: (a) there is a natural barrier to access these large models, and we can get only the output of the model for a given input via some interface (e.g., REST APIs), (b) these large language models constantly evolve and get better with each generation~\cite{GPT2,GPT3}; treating them as black-boxes enables plug-and-play with minimal effort, and (c) finally, domain-specific improvements to large language models (e.g. for Python programming~\cite{PyMT5}, general programming~\cite{Codex}) are rendered complementary to our efforts rather than competitive.

We configure~\jigsaw~with a~\lm~(GPT-3 and \codex, in this work)~of choice to be used as a black-box. We focus on appropriately setting up or priming these models for a given task at hand, characterizing common failure modes of ~\lms{}s for code synthesis, and building components that can overcome such recurring failures. We rely on both program synthesis-based techniques and multi-modal specification to design these components. Our goal is to enable synthesis of syntactically and semantically correct code snippets for a given domain and using feedback from usage to improve the system over time.

The pre-processing module of ~\jigsaw~ contextualizes the input to the black-box language model using heuristic techniques (akin to  recent efforts~\cite{Zhao2021Calibrate,Perez2021TrueFL,promptselection}). A key contribution of our system is in its post-processing module:(a) speeding up the combinatorial search space of API functions and their arguments, and (b) learning and updating a set of transformation or re-write rules to be applied to the erroneous snippets output by~\lms{}s. The post-processing module uses I/O examples to choose the appropriate transformation.

In the rest of the paper, we instantiate \jigsaw~for solving data transformation tasks with Python Pandas API,
which is widely used by data scientists to process tabular data~\cite{PandasAPI}.

\subsection{\lms{}s as black-box}
Consider a typical scenario where the user wants to load and examine the data in a \texttt{csv} file. Pandas uses dataframe objects (two-dimensional tabular representation) to store and process heterogeneous data (commonly named with \textit{df} prefixes in the code, like \textit{df, df1, df2, dfin}). The user invokes \jigsaw~ with a simple natural language description of the intent in a cell of Jupyter Notebook:

\texttt{\%jigsaw -q "Load ./data.csv file"}

The ``magic command'' \texttt{\%jigsaw} invokes the synthesis pipeline with the given query.~\jigsaw~(configured with \gpt) returns the following snippets for the above query:

\begin{lstlisting}
df = pd.read_csv('./data.csv')
csv = pd.read_csv('./data.csv', header=None)
\end{lstlisting}

The user then issues the following query in the session:

\texttt{\%jigsaw -q "Remove substring `Name:' from column `country' of df"}

\jigsaw~produces the following snippets.
\begin{lstlisting}
df['country'] = df['country'].str.replace('Name:', '')
\end{lstlisting}

A key knob in \lms{}s is setting the right context for a given user query. This context is passed as an input to the \lm\ in addition to the user query. To this end, \jigsaw~first prepares the input in the \textit{pre-processing stage} (details in Section~\ref{sec:preprocessing}). The preparation involves assembling a set of ``relevant'' question-answer pairs to inform the~\lm~ of the nature of the input task --- which is converting natural language text to Python code, specifically, Pandas code. With the context selection in the pre-processing stage,~\jigsaw~produces the desired code snippets shown above. In contrast, the GPT-3 model without context selection produces the following incorrect snippet for the above query:\\

\begin{lstlisting}
df = df.country.str.remove('Name:')
\end{lstlisting}

Recent studies, both in the context of natural language understanding~\cite{Zhao2021Calibrate,Perez2021TrueFL} as well as in the programming~\cite{GoogleMNN}~domains, have shown the influence and importance of context selection in the output of ~\lms{}s. Our work provides further evidence that context selection can significantly impact the quality of the code generated for Pandas programming tasks, with two different~\lms{}s, demonstrated in Section~\ref{subsec:ablations}. 

\subsection{Learning to fix recurring failure modes of~\lms{}s}
The core aspect of \jigsaw~system design is incorporating a post-processing phase that involves: (a) characterizing, (b) transforming the (syntactically and/or semantically) erroneous code snippets, and, more importantly, (c) endowing the system with the capability to improve (in terms of accuracy) from feedback as more users interact with it over time. Below, we highlight common classes of errors we observe over two different Pandas programming datasets (created by us, and described in Section~\ref{sec:benchmarks}) using two different~\lms{}s, namely GPT-3 and \codex.\\
\textbf{1. Referencing errors:} We observe that, even with suitable context, \lm{}s can produce incorrect referencing of variable names in otherwise accurate code snippets. \\
\textbf{2. Incorrect arguments:} In some cases,~\lm{}s produce code with the right composition of API functions, but with incorrect arguments. For instance, consider the following invocation:\\
\texttt{\%jigsaw -q "remove all duplicate entries of column 'inputB'"} 
\begin{lstlisting}
dfout = df.drop_duplicates(subset=['inpB']) # PTLM
dfout = df.drop_duplicates(subset=['inpB'],keep=False) # Correct
\end{lstlisting}
\textbf{3. Semantic errors:} A recurring failure mode for the~\lm{}s we have experimented with is that they produce code snippets that are \textit{almost} correct, but the semantics are wrong because of a minor error. We can quantify this via suitable edit distance between the ASTs of the produced and the correct (i.e., intended) code snippets. For instance, consider the following invocation:

\texttt{\%jigsaw -q "Get fourth value from column 'C' in dfin and assign to dfout"}
\begin{lstlisting}
dfout = dfin.ix[3, 'C'] # PTLM
dfout = dfin.loc[3, 'C'] # Correct
\end{lstlisting}

~\jigsaw~employs a post-processing phase that critically relies on the multi-modal specification (I/O examples, in particular) to overcome the aforementioned recurring failures. To this end, we pass the incorrect output code snippet from~\lm~(which can be ascertained with the help of I/O examples in the specification) through a series of components driven by PL-based techniques (details in Section~\ref{sec:postprocessing}). The two key ideas are outlined below. \begin{enumerate}
    \item Using the API functions in the incorrect code snippets produced by~\lm{}, we seed the enumerative search for the right arguments.
    We perform this search efficiently adapting the \ap~framework~\cite{Autopandas} which is an enumerative-search based programming by examples framework built for Pandas API.
    \item The user interface of \jigsaw~enables getting feedback which is then used by our system to learn a set of AST-to-AST transformations using the \prose~synthesis framework~\cite{Prose,Refazer}. The challenge here lies in clustering errors that are \textit{alike} so that a small set of general transformations can be learnt.
\end{enumerate}

\section{Jigsaw Architecture}
\label{sec:architecture}

The architecture of \jigsaw\ is depicted in Figure~\ref{fig:jigsaw-architecture}.
In this section, we describe each module in detail.

\subsection{Pre-trained Language Models}
\label{subsec:PTLM}
We describe Pre-trained Language Models (\lm s) using \gpt\ as an example. \gpt\ stands for "Generative Pre-trained Transformer 3", which is the third version of a large transformer model developed by OpenAI.
\gpt\ is a neural model with 175 billion parameters, trained on a very large corpus consisting of publicly available datasets such as  CommonCrawl~\footnote{https://commoncrawl.org/the-data}, WebRText dataset, two internet-based books corpora, and English Wikipedia.
\gpt\ is a general-purpose model that can be customized to perform a variety of NLP tasks. Such customizations do not involve fine-tuning the ML model for the specific task at hand. Instead, the user of \gpt\ can describe the task using a few examples (on the order of 4-5 examples works usually), and \gpt\ is then able to produce answers for the specific task.
A session with \gpt\ has the form: 
$(Q_1, A_1), (Q_2, A_2), \ldots, (Q_k, A_k), Q$, where $k$ is a small number (typically 4 or 5), the pairs $(Q_i, A_i)$ are question-answer pairs to describe the task we want \gpt\ to perform, and $Q$ is the question for which we seek an answer. 

\begin{figure}
    \begin{lstlisting}[basicstyle=\scriptsize]
    gpt3 = GPT(engine="davinci", temperature=0.5, max_tokens=100)
    # Examples to train a English to French translator
    gpt3.add_example(Example('What is your name?', 'quel est votre nom?'))
    gpt3.add_example(Example('What are you doing?', 'Que faites-vois?'))
    gpt3.add_example(Example('How are you?', 'Comment allex-vous?'))
    
    # Input to the model
    prompt3 = "where are you?"
    output3 = gpt3.submit_request(prompt3)
    # Model output
    output3.choices.text\end{lstlisting}
    \hskip-2.0cm
    \begin{minipage}{1in}
    \begin{verbatim}
    Output: Où êtes-vous?  
    \end{verbatim}
    \end{minipage}
    \vspace{-3ex}
    \caption{English to French translation using \gpt}
    \label{fig:english-french-gpt3}
\end{figure}

For example, if $(Q_i, A_i)$ are such that $Q_i$ are English statements and $A_i$ are corresponding French translations, then \gpt\ becomes an English-French language translator. See session in Figure~\ref{fig:english-french-gpt3}.


Other recent \lm s include Codex~\cite{Codex}, which is OpenAI's recent language model trained specifically on code, and Google's large language model~\cite{GoogleMNN}; these models translate natural language to program. \jigsaw\ uses \lm s to produce Pandas code, given a natural language description of intent, and test cases. Specifically, \jigsaw\ session with \gpt\ has the form: 
 $(N_1, P_1), (N_2,  P_2), \ldots, (N_k, P_k), N$
where $N_i$ is English description of intent,  and $P_i$ is the code snippet we want the \lm\ to produce. We currently do not pass input-output examples to the \lm. Instead, we use these test cases to check and filter the candidate codes produced by the \lm\ during post-processing, or transform the code produced by the \lm\ such that it passes the test cases. 

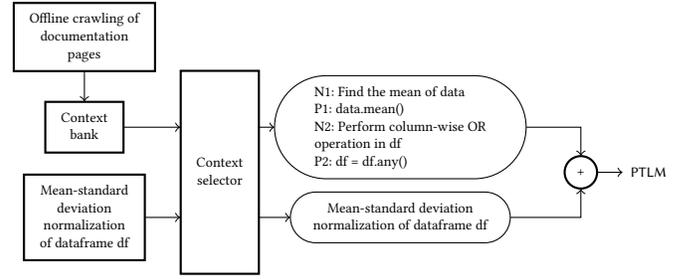
\begin{figure}
\begin{tikzpicture}[every node/.style={inner sep=5, scale=0.6, every node/.style={scale=0.5}}]
\usetikzlibrary{shapes.misc, positioning}
\node[draw,  rectangle, thick, xshift=-5cm] (n0) {\begin{tabular}{c} Offline crawling of \\ documentation \\ pages \end{tabular}} ;
\node[draw,  rectangle, below of=n0, thick, yshift=-1cm] (n1) {\begin{tabular}{c} Context \\ bank \end{tabular}} ;
\node[draw, rectangle, thick, below of=n1, yshift=-1cm] (n2) { \begin{tabular}{c} Mean-standard \\ deviation \\ normalization \\ of dataframe df \end{tabular} } ;
\node[draw, rectangle, thick, right of=n2, xshift=2cm, yshift=1cm, minimum height=4.5cm] (n3) {\begin{tabular}{c} Context \\ selector \end{tabular}} ;
\node[draw, rounded rectangle, right of=n3, xshift=3cm, yshift=1cm] (n4) { \begin{tabular}{l} N1: Find the mean of data\\ P1: data.mean() \\ N2: Perform column-wise OR  \\ operation in df \\ P2: df = df.any() \end{tabular} } ;
\node[draw, rounded rectangle, right of=n3, xshift=3cm, yshift=-1cm] (n5) { \begin{tabular}{c} Mean-standard deviation \\ normalization of dataframe df \end{tabular} } ;
\node[draw, circle, thick, right of=n4, xshift=3cm, yshift=-1cm] (n6) { + } ;
\node[right of=n6, xshift=0.5cm] (n7) { \textbf{\lm} } ;


\path[draw,->] (n0) -- (n1) ;
\path[draw,->] ($ (n1.east) + (0cm, 0cm) $) -- ($ (n3.west) + (0cm, 0.6cm) $) ;
\path[draw,->] ($ (n2.east) + (0cm, 0cm) $) -- ($ (n3.west) + (0cm, -0.6cm) $) ;
\path[draw,->] ($ (n3.east) + (0cm, 0.6cm) $) -- ($ (n4.west) + (0cm, 0.0cm) $) ;
\path[draw,->] ($ (n3.east) + (0cm, -0.6cm) $) -- ($ (n5.west) + (0cm, 0.0cm) $) ;

\path[draw,->] (n4) -| ($ (n6.north) + (0cm, 0cm) $) ;
\path[draw,->] (n5) -| ($ (n6.south) + (0cm, 0cm) $) ;
\path[draw,->] (n6) -- (n7) ;
\end{tikzpicture}
\vspace{-3ex}
\caption{Illustration of Pre-processing step of \jigsaw}
\label{fig:pre-processing}
\end{figure}

 \subsection{Pre-processing}
 \label{sec:preprocessing}
The goal of \jigsaw's pre-processing module is to convert the user intent into a suitable query for the \lm. As mentioned above, \lm s take a sequence of question-answer pairs $(N_1, P_1), (N_2,  P_2),$ $\ldots ,$ $(N_k,P_k)$ as a preamble before we supply the current query. We use the term {\em  context}  to denote this preamble. 
Previous works in natural language processing (NLP) ~\cite{Zhao2021Calibrate,Perez2021TrueFL,promptselection} have shown that the performance of these \lm s is heavily influenced by this context; and the performance improves if the question-answer pairs in context are  similar to the current query $N$. 
Hence, we maintain a  {\em context bank}, of possible question-answer pairs, and then choose  elements of the context bank that are similar to the current query,  and add these to the context. 
\jigsaw\ creates a context bank offline by scraping and annotating examples from API documentation for Pandas, as well as other resources of examples (from tutorials, etc.) that
are used to teach the API. When the user asks a question, the question is fed to a context selector that uses a text similarity metric to pick the most relevant prompts from the context bank (see Figure~\ref{fig:pre-processing}). We study two kinds of similarity metrics for the context selector: 
\begin{inlineenum}
\item tf-idf similarity ~\cite{tfidf} (\tfidf), and
\item transformer similarity ~\cite{sentenceTransformer}  (\transformer)
\end{inlineenum}. The context thus produced is appended with the current query and fed as input to the \lm.

In cases where \jigsaw\ is unable to produce the correct answer, we let users make changes to the incorrect \jigsaw\ code and use such a feedback to enhance the context bank (details in Section~\ref{subsec:learning-feedback}). \lm s also take an input parameter called {\em temperature}. Lower values of temperature result in fewer accurate answers. Higher values result in a less accurate but more diverse set of answers. We report on how we pick temperature values in Section~\ref{sec:experiments}. 

 \subsection{Post-processing}
 \label{sec:postprocessing}
The code snippets produced by the \lm\  vary in accuracy and quality, depending on the natural language sentences used to ask to encode the question, the context bank supplied, the context selection as well as the temperature parameter.
The goal of \jigsaw's post-processing step is to filter and transform the output produced by the \lm\  to produce a correct answer.
Our measure of correctness is that the code produced should pass the I/O examples specified by the user.
In many cases, the code does not parse or fails with an exception. We consider such cases as test failures.
If a non-empty set of candidate solutions produced by the \lm\ satisfies the test cases, then we merely show those code snippets to the user. Our experience is that for about $30$\%-$60$\% of the cases (depending on the \lm\ and the dataset), the \lm\ produces correct outputs.
In the remaining cases,
\jigsaw\ uses the candidate solutions produced by \lm\ as starting points and performs transformations on candidate code snippets using simple program analysis and synthesis techniques to produce correct solutions. We describe the correctness checks and transformations below:
\\
\noindent{\bf Correctness checks}: In cases where we have I/O examples, we run the candidate code snippet starting with each of the specified  inputs, and check if the output produced agrees with the corresponding specified output.  This check can be expanded to include static analyses to check for security vulnerabilities and other errors, as motivated by recent work~\cite{CodexVulnerabilitiesPaper}.
\\
\lstMakeShortInline[columns=fixed]@
\noindent{\bf \nameTransforms}: In some cases, \lm s produce accurate code snippets, but with incorrect variable names. This is often due to the model's bias towards common variable names like @df@ for Pandas dataframes and also because users assume variable referencing to be implicit.
As an example, we find that \gpt\ produces the code snippet @df1.merge(df2)@ when the correct answer is @df2.merge(df1)@.
Since users specify inputs and output variables in the natural language description or in test cases, this post-processing step uses such information from multi-modal inputs, as well as names of variables in scope, by systematically searching over potential variables, and trying possible permutations and combinations of variable names so as to pass the test cases. 
\lstDeleteShortInline@
\\
\noindent{\bf \argTransforms}: 
In some cases, the \lm s produce code snippets with correct method names and method sequences (in case multiple methods need to be invoked in a nested manner or one after another), but with incorrect arguments.
As an example, in response to the query  ``{\tt replace `United States' in `location' by `US' and `3434' in `zip' by `4343'}'',  \codex\ produces:
\begin{lstlisting}
dfout = df.replace({'United States':'US', 3434:4343})
\end{lstlisting}
This snippet invokes the correct method {\tt replace}, but misses the detail in the question that {\tt `United States'} and {\tt 3434} must be replaced with {\tt `US'} and {\tt 4343} \textit{only} when these values are present in the columns {\tt `location'} and {\tt `zip'} respectively. The correct code synthesized by \jigsaw\ for this query is as shown below:
\begin{lstlisting}
dfout = dfin.replace({'location':{'United States':'US'},
        'zip':{3434:4343}})
\end{lstlisting}
Motivated by such cases, this post-processing step 
systematically searches through arguments from an inferred argument space for a given sequence of method/function names.  In order to implement the systematic search over the space of arguments, we adapt the approach used by Autopandas tool~\cite{Autopandas}, with the following modifications. Autopandas uses a Graph Neural Network, that takes I/O examples as input, to choose method names. However, we need a lot of domain-specific data to train such neural networks. In our case, we simply extract the method names from the output of \lm~given the natural language query (which readily scales to programming domains beyond Pandas). The argument space to perform the search is inferred using the natural language text input, the arguments present in the \lm\ output, the column names from the dataframe schema as well as variables in scope.
We extend the generators in Autopandas to consider complex data types such as lists and dictionaries, and we extend the set of APIs considered to include APIs that return Pandas Series types ( one-dimensional labeled arrays capable of holding data of any type) in addition to the ones that return Pandas dataframe types. With these modifications, we find that \jigsaw\ is able to transform several incorrect code snippets produced by the \lm\ to correct code snippets (as shown in Section~\ref{sec:experiments}). 
\\
\noindent{\bf \astTransforms}: 
In some cases, we find that the \lm{} produces code that is almost correct but has a minor error. We also find that such errors are {\em repeatedly} made by the PLTM, and are fixable with suitable AST-to-AST transformations, learned from user interactions with \jigsaw.
As a specific example, we find \gpt\ often misses the bitwise not operator, and produces the code:
\begin{lstlisting}
train = data[data.index.isin(test.index)]}
\end{lstlisting}
instead of the following correct code with the bitwise not operator:  
\begin{lstlisting}
train = data[~data.index.isin(test.index)]}
\end{lstlisting}
As another example, we find that \gpt\ misses paranthesizations, which results in the generated code raising an exception. Specifically, the generated code is:
\begin{lstlisting}
dfout = dfin[dfin['bar']<38|dfin['bar']>60]
\end{lstlisting}
instead of the following code synthesized by \jigsaw\, which is parenthesized correctly:
\begin{lstlisting}
dfout = dfin[(dfin['bar']<38)|(dfin['bar']>60)] 
\end{lstlisting}

Such errors cannot be fixed via variable name transformation or argument transformations. 
\jigsaw\ corrects such errors by learning  re-writing rules as AST-to-AST transformations.
These transformations are applications of production rules from grammar used in BluePencil~\cite{bluepencil} which is used for suggesting code re-factorings. However, it is not possible to learn these rules at the appropriate level of generality from a single example. This generality is necessary so that the missing negation or parenthesizing can be corrected by the learnt transformation, even if the same pattern is repeated with a different set of variables or constants.
To achieve this, we collect data from user interactions, where the user edits the answer produced by \jigsaw\ to produce the correct code. We cluster the data points (i.e., code snippets) so that similar data points are grouped together and we learn a single AST-to-AST transformation that is able to handle all the data points in a cluster. 
Unlike the case of refactoring where users will implicitly hint at clustering of similar edits (by attempting them one after the other), we resort to a greedy heuristics-based clustering algorithm.
This clustering is performed in an online fashion as we get more data points for learning AST-to-AST transformations. For each data point, we decide if the data point is grouped inside an existing cluster or instantiate a new cluster. 
In the former case, we check if the AST-to-AST transformations from the existing cluster can be re-learnt to be more general, and if so, re-learn the transformations. 
In addition, we perturb the data points in each cluster to change variable names and constants, in order to prevent learning transformations that over-fit. 
Together with the above-mentioned clustering and perturbation heuristics, we find that we are able to learn transformations at the appropriate level of generality (Section ~\ref{subsec:transformations}). Code has well-defined structure, usually represented as abstract syntax trees. We take advantage of this structure while learning these AST-to-AST tree transformations. 
We use the PROSE program synthesis system~\cite{Prose,Refazer}~to learn the transformations from a cluster of incorrect-correct code snippets.
While \jigsaw\ currently works only on Python code, the post-processing step works at the level of ASTs, and can be made to work across programming languages as well. For instance, in \cite{bluepencil} the same tools are deployed to learn  non-trivial code refactoring in C\#, SQL, Markdown, and spreadsheets.

We refer to \argTransforms\ and \astTransforms\ together as \semrep\ in experiments (Section~\ref{sec:experiments}).

\subsection{Learning from user feedback}
\label{subsec:learning-feedback}
The user interface of Jigsaw (integrated into the Jupyter notebook) is designed to let users submit correct code in cases where Jigsaw is incorrect. \jigsaw\ can be improved by assimilating user feedback. Specifically, we design techniques for updating context-bank in the pre-processing module and \astTransforms\ in the post-processing steps, as more users interact with \jigsaw.\\
{\bf Updating context bank:} The procedure for updating context bank with user queries is given in Algorithm~\ref{algo:updateBank}. We first check whether \jigsaw\ already found a correct solution for the given (new) query $N$, thus giving us some confidence about its correctness. Otherwise, we check if any of the solutions generated by \jigsaw\ is ``close'' to some correct code (determined by the standard edit distance on strings $\mathbf{d}_{\algoedit}$ and a chosen threshold $\epsilon_{\algocode}$). If either of the two conditions is satisfied, we add the new query to our context bank while additionally ensuring that a similar query already does not exist (via \tfidf\ based distance $\mathbf{d}_{\tfidf}$ and a threshold $\epsilon_{\algobank}$). 
\begin{algorithm}
	\begin{algorithmic}
        \item \textbf{Inputs: } 
        
        \indent Context Bank : $\mathC = \{(N_1\text{, }P_1),(N_2\text{, }P_2),\dots,(N_{|\mathC|}\text{, }P_{|\mathC|})\}$, 
        
        \indent New query and feedback (code snippet):  $N, P$
        \item \textbf{Output: } Updated Context bank $\mathC$
        \item Let output = \jigsaw($N$, $\mathC$)
        
        \item If $\min_i \mathbf{d}_{\algoedit}(\text{output}_i, P) > \epsilon_{\algocode}$
        
            return $\mathC$

        \item If $\max_i \mathbf{d}_{\tfidf}(N, N_i) < \epsilon_{\algobank}$
            return $\mathC$
        
        \item return $\mathC \cup \{(N, P)\}$ 
 	\end{algorithmic}
	\caption{Updating context bank}
	\label{algo:updateBank}
\end{algorithm}
With more usage, we grow the context bank and try to cover different styles of user queries, which in turn helps relevant context selection. \\
\textbf{Updating transformations:}
For every query paired with correct code snippet(s), we select all incorrect codes suggested by \lm\ within some small edit distance of a correct code. The \astTransforms\ learning sub-module performs clustering (with perturbations) on the selected code snippets as discussed in the above subsection, and updates the set of transformations.

While the above pre-processing and post-processing steps were designed in the context of the Python Pandas API, we believe that ideas such as context selection, correctness checking, and transformations are general and that it is possible to design pre-processing and post-processing steps in a generic manner that can work across languages and APIs.  For each API, specific transformation rules can be learnt from usage data generated by users of that API.

\section{Datasets}
\label{sec:benchmarks}
We perform our experiments on two different datasets 
\footnote{The datasets can be found at \href{https://github.com/microsoft/JigsawDataset}{https://github.com/microsoft/JigsawDataset}}.
\subsection{PandasEval dataset \webapp}
This dataset consists of $68$ Python Pandas tasks.
Each task can be solved using a single line of code by composing at most 2-3 Pandas functions; sometimes followed by assigning variables. This dataset was created by authors of this paper by going through queries in online forums like StackOverflow. An example task from this dataset is ``For every row in df1, update `common' column to True if value in column `A' of df1 also lies in column `B' of df2". 

\subsection{Hackathon dataset \hackathon }
\label{subsec:hackdata}
\begin{figure}[h]
\includegraphics[keepaspectratio, width=0.9\linewidth]{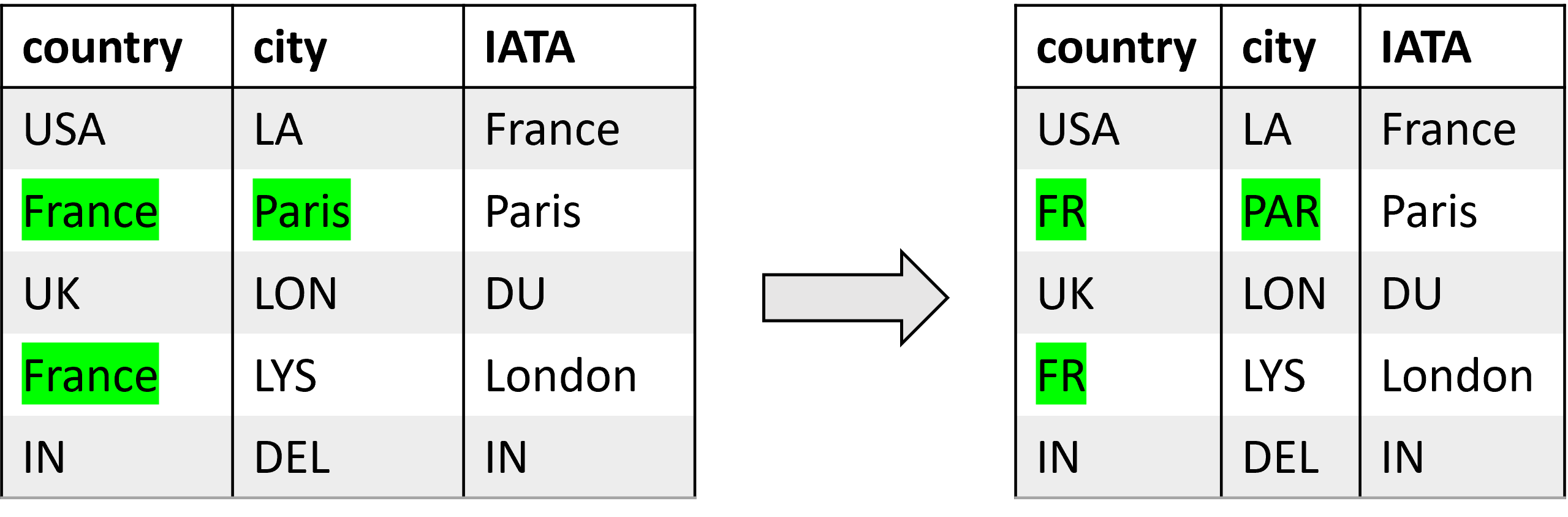}
\caption{Example task, part of the dataset~\hackathon, as presented to the user during the Hackathon session.}
\label{fig:example-task1}
\vspace{-0.5cm}
\end{figure}

This dataset consists of $21$ Pandas tasks; each task can be solved by composing at most 2-3 Pandas functions, possibly followed by assigning variables, as in the~\webapp~ dataset. We posed these tasks as illustrations, in a hackathon we conducted with $25$ users over $2$ different sessions. Table~\ref{tab:profile} presents self-reported proficiency of the users in Python and Pandas. An example illustration that shows the intent of a task is given in Figure~\ref{fig:example-task1}. Users were asked to read such pictorial illustrations and come up with their own natural language (English) query constructions for each task. We then collected the queries written by the users, clustered, and annotated them to produce the~\hackathon\ dataset comprising of a total $725$ unique queries constructions. The task corresponding to the illustration in Figure~\ref{fig:example-task1}, from the dataset~\hackathon, is shown below. Here \texttt{dfin} and \texttt{dfout} refer to the dataframes in Figure~\ref{fig:example-task1}. 

\begin{table}[]
    \centering
    \begin{tabular}{|c|c|c|c|}
    \hline
         & Beginner & Intermediate & Advanced \\
         \hline
        Python & 1 & 21 & 3 \\
        Pandas & 17 & 8 & 0 \\
    \hline
    \end{tabular}
    \caption{Proficiency of participants from \hackathonOffline\ dataset}
    \label{tab:profile}
    \vspace{-25pt}
\end{table}

We note that while users provided precise and clear natural language queries in many cases, they also came up with imprecise and incorrect formulations in some cases. For instance, in the spec shown above, the query provided by \texttt{user1} is correct, whereas the one provided by \texttt{user2} is incorrect because the word ``{\em France}'' is present in the ``{\em IATA}'' column as well; Figure~\ref{fig:example-task1} conveys that only the ``{\em country}'' column needs to change, and not the ``{\em IATA}'' column. 

Since such queries were created by users interacting with the system, and users tend to make mistakes, it is useful to have such variations in the dataset. While curating the dataset, we removed natural language queries that were clearly incorrect, and retained queries that were imprecise and partially correct.

\begin{lstlisting}[language=json,firstnumber=1,caption=Example json for a task in PandasEval2]
"task_8": {
		"queries": [
    ["replace 'France' with 'FR' in 'country' column and 'Paris' with 'PAR' in 'city' column", "user1"],
    ["In dataframe dfin, replace cells having 'France' to 'FR' and cells having 'Paris' to 'PR'", "user2"]
		],
		"IO": [{
			"inputs": "dfin",
			"output": "dfout"
		}]
	}
\end{lstlisting}

As mentioned earlier, we conducted the hackathon over two sessions. We use \hackathonOnlineOne\ to denote the dataset generated from user queries from the first session, and \hackathonOnlineTwo\ to denote the dataset generated from user queries from the second session. 
For each of the $21$ tasks, we created semantic variations (e.g. changing constants, API arguments) of the same task. Consequently, users in the second session (\hackathonOnlineTwo) saw different variants of the tasks when compared to users in the first session (\hackathonOnlineOne). Specifically, $3$ tasks were exactly the same, $9$ had differences in constants and $9$ had changes in arguments. We introduced these variants in order to study if \jigsaw\ can learn from usage in the first session to improve user experience in the second session (see Section 5.2).
We use ~\hackathonOffline~to denote the union of \hackathonOnlineOne\ and \hackathonOnlineTwo.


\section{Experiments}
\label{sec:experiments}
\lstMakeShortInline[columns=fixed]@

\begin{table*}[th]
\begin{tabular}{lllll|lll}
\toprule
     &     & \multicolumn{3}{c}{\webapp} & \multicolumn{3}{c}{\textbf{\hackathon}} \\
     &     &          \lm &       Variable Name &           \semrep &          \lm &       Variable Name &           \semrep \\
\midrule
\multirow{2}{*}{\gpt} 
& \no &  $30.9\pm1.2$ &  $38.2\pm2.4$ &  $44.6\pm3.9$ &   $8.9\pm0.6$ &  $24.8\pm0.9$ &   \textbf{$33.6\pm0.5$} \\
&\transformer &  $33.8\pm2.4$ &  $41.7\pm2.5$ &  $47.1\pm2.1$ &   $6.6\pm0.2$ &  $24.3\pm0.8$ &  \textbf{$35.1\pm0.7$} \\

\cline{1-8} 
\multirow{2}{*}{\codex}      
& \no &  $45.6\pm1.2$ &  $54.9\pm0.7$ &  $59.8\pm3.5$ &  $26.8\pm1.2$ &  $51.0\pm0.6$ &  \textbf{$56.8\pm0.3$} \\
& \transformer &  $52.0\pm0.7$ &  $63.7\pm0.7$ &  $66.7\pm0.7$ &  $31.2\pm0.2$ &  $67.5\pm0.5$ &  \textbf{$72.2\pm0.5$} \\

\bottomrule
\end{tabular}
\caption{Performance (mean accuracy $\pm$ std. deviation) after different stages of the \jigsaw\ pipeline on \webapp\ and \hackathon\ datasets. \jigsaw~post-processing steps significantly improves upon \lm{}s irrespective of context selection strategy. Pre-processing clearly benefits, comparing rows 1 vs 2, and 3 vs 4. }
\label{table:offline_evaluation}
\vspace{-0.9cm}
\end{table*}

We evaluate Jigsaw on the two datasets introduced in Section~\ref{sec:benchmarks}, with emphasis on the following questions:
\begin{inlineenum}
\item How accurate is the~\jigsaw~system compared to the black-box~\lm{}s and other code synthesis methods?
\item What is the utility and applicability of the individual~\jigsaw~ components (in the pre-processing and post-processing modules)?
\item Can these components benefit from feedback over time as more users interact with the system?
\end{inlineenum}  

For the first question, we evaluate \jigsaw\ in an offline setting, i.e., without learning from any feedback (in Sections~\ref{subsec:offline}); and present comparisons against the state-of-the-art \ap~framework, which generates Pandas snippets using only I/O example (in Section~\ref{subsec:AP}). For the second question, we perform a temporal study on the \hackathon\ dataset (in Section~\ref{subsec:temporal}), where we leverage user feedback from the first hackathon session to update the system and measure the performance improvement in the second session. We also perform ablation studies (in Section~\ref{subsec:ablations}) pertaining to our context selection sub-module. We end with a preliminary evaluation of Jigsaw on tasks pertaining to the TensforFlow API~(in Section~\ref{subsec:beyond}).

We consider \textbf{accuracy} as our primary evaluation metric, i.e., fraction of specifications in the dataset for which a {\em correct} program was synthesized.
We define a program as correct if it satisfies the given I/O examples, and additionally passes a manual inspection of whether the synthesized code meets the intent of the natural language description. The manual inspection helps us reject programs that satisfy the I/O examples by overfitting on them and violate the general intent of the natural language descriptions.
Note that there is inherent randomness in the output of the~\lm{}s, so we run every evaluation three times and report the mean accuracy (\%) and standard deviation (over the runs). In some cases, we also present \textbf{task completion} metric which is the percentage of tasks correctly solved by a user (regardless of the number of queries used to solve a task) interacting with the system. Furthermore, in every case, we present the best accuracy obtained by varying the temperature parameter of~\lm~$\in \{0, 0.2, 0.4, 0.6\}$.

\subsection{Offline evaluation}
\label{subsec:offline}
In Table~\ref{table:offline_evaluation}, we present the performance of~\jigsaw~on \webapp\ and \hackathon\ datasets, with \gpt\ and \codex~as black-box ~\lm{}s. The second column of the table indicates the context selection strategy for the~\lm. For this study, we consider \no~(no tailored context provided for the user query; we use a default context: ``@import pandas as pd@''), and \transformer\ (Transformer similarity based context selection, discussed in Section~\ref{sec:architecture}) with number of context prompts fixed as $4$. Each cell in the table gives the accuracy metric with mean and standard deviation as defined above. For~\jigsaw, the column titled \nameTransform\ indicates the performance of the system using only this part of the post-processing module; and the column titled \semrep\ indicates the performance of the system in its entirety, i.e., running \nameTransforms\ followed by \semrep\ (\argTransforms\ and \astTransforms). 

Comparing \lm~and~\semrep~columns, it is evident that~\jigsaw~ improves upon the black-box~\lm{}s, in terms of accuracy, by $15$\%-$40$\% irrespective of the context selection strategy, on both the datasets as well as on both the~\lm{}s. These results underscore the utility of program analysis-based augmentation of large language models. 

Next, from Table~\ref{table:offline_evaluation}, we find that providing useful context for the language model along with the query significantly improves upon not providing any context (comparing rows 1 vs 2, and rows 3 vs 4), across the datasets and~\lm{}s. It is clear that \lm\ with \transformer\ context is better than \no\ by a margin $\sim5$\% (without post-processing) and up to $15$\% with post-processing for \codex\ on the two datasets. For \gpt{}, \transformer\ context is significantly better than \no\ on the \hackathon\ dataset and on \webapp, the numbers are statistically insignificant. 
\lm{}s require some initial context in the form of examples to characterize the task to be solved, and these results underscore the importance of having a pre-processing module in~\jigsaw. 

Finally, from Table~\ref{table:offline_evaluation}, we also observe the effectiveness of the individual post-processing modules of~\jigsaw, as discussed below. Note that, for these results, we seed our \astTransforms\ using a small dataset collected from StackOverflow questions. Later, in Section~\ref{subsec:temporal}, we show that these numbers can be significantly improved by learning transformations from usage over time.\\
{\bf\nameTransforms:} \lm{}s make variable referencing errors (as noted in Section~\ref{sec:overview}) because of its implicit bias towards common dataframe names such as \texttt{df, df1, df2, dfout} and also because users tend to not specify variables explicitly in their queries. We find that this simple post-processing module gives an improvement of $10$\%-$30$\% for \codex~and $10$\%-$15$\% for \gpt.\\
{\bf\semrep:} We see that the semantic repair post-processing module improves absolute performance of \codex\ by $\sim5$\% and of \gpt\ by $6$\%-$11$\%. This underscores the significance of using program analysis techniques to augment language models that do not have inherent understanding of code semantics. Recall (from Section~\ref{sec:architecture}) that \semrep~ consists of \argTransforms\ and \astTransforms\ sub-modules.  We find that, using just the ~\argTransforms~(without~\astTransforms), improves absolute performance of the system by $5$\%-$9$\% and $3$\%-$5$\% for \gpt\ and \codex\ respectively (not shown in Table~\ref{table:offline_evaluation}). Similarly, using \astTransforms\ alone (without~\argTransforms), we obtain improvement of up to $3.5$\% for \gpt\ and $1.3$\% for \codex~ (not shown in Table~\ref{table:offline_evaluation}). 

We find that our post-processing steps are reasonably fast; time taken by \jigsaw\ is primarily bottle-necked by the inference times of \lm\ APIs. Specifically, on average getting output from \codex\ takes $\sim7$ seconds while our post-processing module takes $<3$ seconds. Similarly, on average, \gpt\ takes $30$-$40$ seconds for different context sizes while post-processing finishes in $<10$ seconds.


\vspace{-0.4cm}
\subsection{Temporal evaluation}
\label{subsec:temporal}
In this section, we evaluate \jigsaw\ on its ability to learn and improve with user feedback. We perform this evaluation on the \hackathon\ dataset. Recall that the hackathon was organized over two separate sessions; so, we use the submissions and feedback for tasks in the first session, corresponding to the \hackathonOnlineOne~dataset, to (a) update our context bank, (b) learn \astTransforms, and (c) evaluate~\jigsaw~on the \hackathonOnlineTwo~dataset consisting of variants of the tasks in \hackathonOnlineOne~(as described in Section~ \ref{subsec:hackdata}).\\
\noindent {\bf Updating pre-processing module:} We follow Algorithm~\ref{algo:updateBank} (with $\epsilon_{\algocode}=25,  \epsilon_{\algobank}=0.15$) and filter out badly written queries from the first session users. Note that, for the 3 tasks that are identical in the two sessions, we do not make any updates to the context bank. We denote our seeded context bank that was used in the first session (containing $243$ question-answer pairs) with \csone\ and the updated context bank updated resulting from Algorithm~\ref{algo:updateBank} with \cstwo\, containing $371$ (243 seeded + 128 new) question-answer pairs.

\noindent {\bf Updating post-processing module:} We follow the procedure described in Section~\ref{subsec:learning-feedback} to learn AST-to-AST transformations from session one data along with seeded data. We use \tsone\ to denote transformations seeded during session one and \tstwo\ to denote the new/updated transformations learned from session one data.


We compare the performance of \jigsaw~on \hackathonOnlineOne~(with \csone~and \tsone) against~\hackathonOnlineTwo~(with baseline \csone~and \tsone, as well as with the updated context bank \cstwo~and transformations~\tstwo) in Table~\ref{tab:temporal}. Each cell in the table is the mean accuracy and standard deviation on the corresponding dataset. Two observations are in order. \\
\textbf{(1) Learning helps improve~\jigsaw.} It is evident that the performance of~\jigsaw~on the \hackathonOnlineTwo\ dataset with the default \csone-\tsone\ setting (column 3) is significantly lower than that of the updated \cstwo-\tstwo~setting (column 4) for both the~\lm{}s. Accuracy of the system with \gpt\ improves by over 30\% due to the updated modules; even with \codex, which already performed quite well on all datasets, we still improve by $\sim15$\% with updates. \\
\textbf{(2) Second session was in general more challenging.} We also observe that the performance on \hackathonOnlineTwo\ with the default \csone-\tsone\ setting (column 3) is significantly lower than that on \hackathonOnlineOne\ with the same setting (column 2). This is because in general the second session was more challenging; partly due to the higher percentage of queries on difficult tasks, and the semantic differences in tasks across the two sessions. But when we use the updated the context and transformations banks, we find a drastic improvement in the performance on \hackathonOnlineTwo, as highlighted in (1) above. This illustrates that \jigsaw\ has the ability to improve from user feedback, regardless of the \lm\ used.

Finally, we also look at the task completion metric (described in the beginning of Section~\ref{sec:experiments}), to assess how the performance gains of learning from feedback translated to user experience during the hackathon. In session one, users we able to solve only $71$\% of the tasks on average; however, in session two, users were able to solve $82$\% of the tasks on average, thus making the experience of the \jigsaw\ system more productive with the updates.

\out{
\naman{We hypothesize that context bank and learned transformations will converge \textbf{quickly? compared to what?} since we are doing synthesizing  API level snippets. \textbf{maybe in discussion?} \textbf{TODO : check with rahul about ``verify with evaluation''}}
}



\begin{table}
\begin{tabular}{ll|ll}
\toprule
{} & \hackathonOnlineOne & \multicolumn{2}{l}{\hackathonOnlineTwo} \\
 &      \csone-\tsone &      \csone-\tsone & \cstwo-\tstwo \\

\midrule
\gpt   &       $45.9\pm0.4$ &       $35.1\pm0.8$ &  $67.2\pm0.3$ \\
\codex &       $75.1\pm0.5$ &       $69.0\pm0.7$ &  $84.4\pm0.8$ \\
\bottomrule
\end{tabular}
\caption{Performance (mean accuracy $\pm$ std. deviation) of ~\jigsaw~without (\csone-\tsone) and with (\cstwo-\tstwo) learning context bank and transformations from user feedback on the \hackathonOffline~dataset. Learning helps improve accuracy significantly, comparing columns 3 and 4.}
\label{tab:temporal}
\vspace{-1.1cm}
\end{table}

\begin{table*}
    \begin{tabular}{lll}
        \toprule
        Code Before & Code After & Semantic Explanation\\
        \midrule
		@out=data[data.index.isin(test.index)]@ & 
		@out=data[@\HighlightFrom@~@\HighlightTo @data.index.isin(test.index)]@  &
		Adding bitwise not inside subscript \\

		@df=df[df['foo']>70|df['foo']<34]@ &
		@df=df[@\HighlightFrom @(@\HighlightTo @df['foo']>70@\HighlightFrom @)@\HighlightTo @|@\HighlightFrom @(@\HighlightTo @df['foo']<34@\HighlightFrom @)@\HighlightTo @]@ &
		Parenthesizing mistake\tablefootnote{Interestingly, this missing parenthesis mistake is quite common and frequented even by humans! See blog post~\cite{parenthesis-blog} and StackOverflow question~\cite{parenthesis-so}.} 
 \\

		@out=df.iloc[0,"HP"]@ &
		@out=df.@\HighlightFrom @loc@\HighlightTo @[0,"HP"]@ &
		Changing @iloc@ to @loc@ \\

		@dfout=df1.append(df2@\HighlightFrom@,ignore_index=True@\HighlightTo@)@ &
		@dfout=df1.append(df2)@ &
		Dropping the last keyword argument \\

		@dfout=dfin.duplicated()@ &
		@dfout=dfin.duplicated()@\HighlightFrom@.sum()@\HighlightTo  &
		Computing sum of series using @.sum()@ \\

		@train=data.drop(test)@ &
		@train=data.drop(test@\HighlightFrom@.index@\HighlightTo@)@ &
		Adding @.index@ in first argument (of drop) \\

		@dfin=dfin["A"].rolling(window=3).mean()@ &
		@dfin@\HighlightFrom@["A"]@\HighlightTo@=dfin["A"].rolling(3).mean()@ &
		Reassign back to the column \\

		@dfout=dfin[(x<40)|(y>53)&(z==4)]@ &
		@dfout=dfin[@\HighlightFrom@(@\HighlightTo@(x<40)|(y>53)@\HighlightFrom@)@\HighlightTo@&(z==4)]@ &
		Giving precedence to bitwise-or \\
    
        \bottomrule

    \end{tabular}
    \caption{Applications (Code After) of learned transformations on code snippets produced by~\lm{} (Code Before).}
    \label{tab:transforms-semantics}
    \vspace{-0.7cm}
\end{table*}

\subsubsection{Analyzing learned \astTransforms}
\label{subsec:transformations}
We present some of the learned AST-to-AST transformations applied to code snippets produced by GPT-3 in  Table~\ref{tab:transforms-semantics}. The transformations were learned using the clustering and perturbing technique outlined in Section~\ref{sec:postprocessing}.
We see that the code fixes are interpretable and they solve common semantic problems in the outputs of~\lm{}s. Please refer to supplementary material for details of the precise AST-to-AST transformations learnt corresponding to first two rows of Table~\ref{tab:transforms-semantics}.

For instance, consider the rule implied in the first row of Table~\ref{tab:transforms-semantics}, which is of inserting @~@ (bitwise not operator) inside subscript. This transformation, learned using the cluster of \textit{diverse} code snippets in Listing~\ref{lst:bit-not}, is fairly general (this is one of the clusters obtained by running our clustering technique on seeded and session one data). On the other hand, consider the last row of Table~\ref{tab:transforms-semantics}, which was learned using the cluster of code snippets in Listing~\ref{lst:precedence}. Since the clustered snippets follow a similar structure, the learned transformation works only when a new snippet has exactly the same logical conditional operators in the specific order. Thus, the quality of the learned transformations depends on the quality of the clustering and of the code snippets themselves, and we expect that more usage data positively influences the overall quality.

\begin{lstlisting}[label={lst:bit-not},caption=Cluster of code snippets from two different tasks that yields the Bitwise-Not transformation in Table~\ref{tab:transforms-semantics}.]
# Task-1
dfout = df.loc[df.isnull().any(axis=1), :] #incorrect
dfout = df.loc[~df.isnull().any(axis=1)] #correct
# Task-2
df_p = df_p.loc[df_per["Name"].str.contains("Ch")] #incorrect
df_p = df_p.loc[~df_per["Name"].str.contains("Ch")] #correct
\end{lstlisting}

\begin{lstlisting}[float=*,label={lst:precedence},caption=Cluster of code snippets from two different tasks that yields the precedence transformation in Table~\ref{tab:transforms-semantics}.]
#Task-1
dfout = dfin[(dfin["gamma"]<40)|(dfin["gamma"]>53)&(dfin["alpha"]==4)] # incorrect
dfout = dfin[((dfin["gamma"]<40)|(dfin["gamma"]>53))&(dfin["alpha"]==4)] # correct
#Task-2
dfout_per = dfin_per.loc[(dfin_per["alpha"]<140)|(dfin_per["alpha"]>159)&(dfin_per["beta"]==103)] # incorrect
dfout_per = dfin_per.loc[((dfin_per["alpha"]<140)|(dfin_per["alpha"]>159))&(dfin_per["beta"]==103)] # correct
\end{lstlisting}

\vspace{-0.2cm}

\subsection{Comparison to \ap}
\label{subsec:AP}
\ap~(\apsmall)~\cite{Autopandas} is a Pandas program synthesis engine capable of generating programs with two or three Pandas functions. It uses \textit{generators} for enumerating over the Pandas API and guides the search with the help of Graph Neural Networks (GNNs) which operate on the input-output (I/O) dataframe(s) and returns the most likely function sequences and arguments. 

In contrast, we make use of multi-modal specification (both natural language query and I/O examples). Programming by examples is known to have ambiguous under-specifications~\cite{Prose,gulwani2016programming}. 
From our experience this issue is exacerbated for large APIs that provide multiple ways for achieving similar functionalities. For instance, consider the specification in Figure~\ref{fig:problem-instance}. If we only consider the I/O example for the given task, we can find many trivial solutions that just drop or select certain rows of dataframe. 

We evaluate \apsmall\ on our \webapp\ and \hackathonOffline\ datasets. As discussed in Section~\ref{sec:architecture}, \apsmall\ does not support series operations, column assignments and dictionary or list generators, many of which are necessary in Pandas workflows. So, out of $68$ tasks in the \webapp\ dataset and $21$ tasks in the \hackathonOffline\ dataset, only $7$ and $20$ are covered by the \ap\ framework respectively. Hence, we compare \jigsaw\ (instantiated with the \codex\ \lm) against \apsmall\ only on these $27$ tasks and use a timeout of $3$ minutes. In the first row of Table~\ref{tab:baseline}, we see that \jigsaw\ clearly outperforms \ap\ even in the restricted subset solvable by \apsmall. This is because $16$ of the $27$ tasks are under-specified if only I/O examples are used and \apsmall\ returns over-fitting solution on many of these tasks; this highlights the necessity of multi-modality. 


We also run \jigsaw\ on the $\apsmall$ dataset~\cite{Autopandas}, where all tasks are supported by \ap\ and I/O examples are sufficient. 
This dataset has been sourced from StackOverflow posts.
Since \jigsaw\ uses text as the primary input, we add natural language descriptions in these posts for querying \codex. 
The results are in the second row of Table~\ref{tab:baseline};
while \codex\ alone is inferior to \apsmall, \jigsaw~(with \codex) performs better than \apsmall. 


\begin{table}[]
    \begin{tabular}{c c c c}
    \toprule
        & \ap~\cite{Autopandas} & \lm & \jigsaw \\
        \midrule
    Subset of \jigsaw\ datasets & 16/27 & 20/27 & 23/27 \\ \ap\ dataset & 17/26 & 15/26 & 19/26 \\
        \bottomrule
    \end{tabular}
    \caption{Number of tasks solved by \jigsaw\ and \apsmall\ on a subset of our dataset supported by \apsmall\ and their dataset.}
    \label{tab:baseline}
    \vspace{-0.8cm}
\end{table}

\subsection{Ablation study}
\label{subsec:ablations}
In both the offline and temporal evaluations presented in the previous subsections, we fixed the number of context prompts to $4$ and \transformer\ as the context selector in the pre-processing module. In this ablation study, we ask if the performance of~\jigsaw~is sensitive to these choices, and  provide justification for the same. All experiments in this section are carried out with the same setting as that of Section~\ref{subsec:offline}.

Table \ref{table:ablation-context-selector} compares the performance of~\jigsaw~ with two different context selection strategies, namely, \tfidf\ and \transformer. We find that the transformer context selector is slightly better, but more importantly, that the performance of~\jigsaw~is not sensitive to the selection strategy. Table \ref{table:ablation-num-prompts} compares the performance of \jigsaw\ with different number of context prompt examples, i.e., 1, 4, and 8. Our experiments show that while there isn't a significant difference between the performances of $4$ prompts vs. $8$ prompts, both perform better than using just $1$ prompt. Again,~\jigsaw~is relatively robust to these choices.

Finally, note that all variations of these choices, for the number of prompts as well as the selection strategy, outperform the \no\ setting (see Table~\ref{table:offline_evaluation}); this further underscores the utility of the pre-processing module. 

\begin{table}

\begin{tabular}{llrr}
\toprule
     & Context &       \webapp &    \hackathon 
     \\

\midrule
\multirow{2}{*}{\gpt} & \tfidf &                  $46.5\pm4.8$ &                     $32.4\pm0.5$ 
\\
     & \transformer &                  $47.1\pm2.1$ &                     $35.1\pm0.7$ 
     \\
\cline{1-4}
\multirow{2}{*}{\codex} & \tfidf &                  $69.1\pm2.4$ &                     $70.1\pm0.1$ 
\\
     & \transformer &                  $66.7\pm0.7$ &                     $72.2\pm0.5$ 
     \\

\bottomrule
\end{tabular}

\caption{Ablation study: Performance of~\jigsaw~with two context selection strategies.}
\label{table:ablation-context-selector}
\vspace{-0.8cm}
\end{table}

\begin{table}
\begin{tabular}{lcrr}
\toprule
     & \# Prompts &       \webapp &    \hackathon \\

\midrule
\multirow{3}{*}{\gpt} & 1 &                  $47.5\pm1.8$ &                     $34.9\pm0.9$ \\
     & 4 &                  $47.1\pm2.1$ &                     $35.1\pm0.7$ \\
     & 8 &                  $48.0\pm2.5$ &                     $32.9\pm0.6$ \\

\cline{1-4}

\multirow{3}{*}{\codex} & 1 &                  $62.3\pm0.7$ &                     $71.8\pm0.5$ \\
     & 4 &                  $66.7\pm0.7$ &                     $72.2\pm0.5$ \\
     & 8 &                  $66.2\pm1.2$ &                     $72.4\pm0.9$\\
\bottomrule
\end{tabular}

\caption{Ablation study: Performance of~\jigsaw~with different number of context prompts.}
\label{table:ablation-num-prompts}
\vspace{-0.8cm}
\end{table}

\subsection{Beyond Pandas}
\label{subsec:beyond}
To test the generality of Jigsaw, we did a preliminary evaluation with $25$ TensorFlow~\cite{TensorFlow}~tasks sourced from TF-coder~\cite{tfcoder} and online forums like StackOverflow. We setup the pre-processing module of \jigsaw~ similar to the offline evaluation, by creating a context bank of $25$ prompts from documentation pages. We reuse the \nameTransform~module and do a \textit{what-if} analysis for argument and tree transformations manually.
\begin{table}[]
    \begin{tabular}{c c c}
    \toprule
        \lm & Variable Name & Semantic Repair \\
        \midrule
        8/25 & 15/25 & 19/25 \\
        \bottomrule
    \end{tabular}
    \caption{Preliminary results of \jigsaw\ with TensforFlow API.}
    \label{tab:tf}
    \vspace{-0.9cm}
\end{table}
Table~\ref{tab:tf}~shows the performance of Jigsaw on the TensorFlow dataset. As seen from the table, Codex alone is able to solve only $8$ of the $25$ tasks, variable transformation improves the performance to $15$ tasks. We manually compare the code outputs to the expected output, to check if argument and tree transformations can be learnt. Based on this analysis, we find that Semantic Repair can potentially improve the performance to $19$ tasks. We show some examples below.
For the query ``Given a tensor in1, replace all instances of 1 with 0'', \lm\ outputs the following:
\begin{lstlisting}
tf.where(x == 1, 0, x)
\end{lstlisting}
The correct code for this query, synthesized by Jigsaw using variable transformation, is shown below:
\begin{lstlisting}
tf.where(in1 == 1, 0, in1)
\end{lstlisting}
For the query ``Given a tensor in1 and a tensor of indices ind, get the sum of elements present at indices in ind from tensor in1.
'', the \lm\ outputs the following incorrect code:
\begin{lstlisting}
tf.gather(in1, ind)
\end{lstlisting}
The correct code, shown below, can be synthesized by Jigsaw with a learnt AST-to-AST transformation, if sufficient data points are collected from usage: \begin{lstlisting}
tf.reduce_sum(tf.gather(in1, ind))
\end{lstlisting}
In summary, this shows that the proposed pre-processing and post-processing modules are useful, and can be generalized to other libraries and programming languages as well.
\out
{
\subsection{Query variations }

\begin{verbatim}
    dfin = dfin.loc[dfin.ftr2 < 5,"ftr2"] = 5
    to
    dfin.loc[dfin.ftr2 < 5,"ftr2"] = 5
\end{verbatim}

}
\lstDeleteShortInline@
 
\section{Threats to Validity}
Our data sets have been created by manually inspecting internet forums like StackOverflow. We tried to cover the common programming patterns in Pandas. 
However, they are not representative of all Pandas programs in the wild.  

We designed the \hackathonOnlineOne\ and \hackathonOnlineTwo\ datasets by collecting data from two sessions of a hackathon, as a proxy for the real-world setting, where large software teams are working on the same project with similar tasks, allowing \jigsaw\ to learn and improve over time.
We varied the tasks between the two sessions, so as to simulate variants of tasks. However, the variations we introduced may not representative of variations of tasks in the real world.
Our study had only 25 participants; evaluating whether the productivity of developers is enhanced in a statistically significant manner in a large scale deployment of Jigsaw is beyond the scope of this paper.

When comparing Jigsaw to \ap, Jigsaw takes as input both the natural language description in the StackOverflow posts and the I/O examples in the posts, while \ap\ only takes the I/O examples as inputs. Hence, Jigsaw has more information about the tasks than \ap. Jigsaw takes less than a minute per task and we use a timeout of three minutes for \ap. Although higher timeouts might improve the performance of \ap\ (10-15 minutes~\cite{Autopandas}), they are not compatible with the interactive user experience that we are aiming for. Whether \ap\ solved a task correctly or not is 
determined by manual inspection and is susceptible to human errors.

\section{Related work}
\label{sec:related}
The literature on using machine learning for program synthesis is vast~\cite{devlin17,balog17,wosuk18,raychev14,sg15,menon13,parisotto17,kalyan18,feng18} and we restrict to works which are closest to Jigsaw (synthesizing code for large APIs using large models and multi-modal specifications).
These works can be classified into the following categories:
1) designed for large APIs but do not use large models, 2) based purely on large models with no multimodal specification, and 3) multimodal synthesis for small APIs. Details follow:
\begin{enumerate}[wide, labelwidth=!, labelindent=0pt]
    \item The TDE~\cite{tde} system for Java relies on rich type information (which is absent in Pandas) and fails to generate argument combinations that are absent from its corpus.
    AutoPandas~\cite{Autopandas} generates Pandas code exclusively from input-output (I/O) examples using a combination of GNNs, which predict  function sequences, and enumerative search. TF-coder~\cite{tfcoder} uses both natural language descriptions and I/O examples to generate TensorFlow code. Both of them use small specific models (as opposed to large generic models like GPT-3)  and lack mechanisms to incorporate user feedback.
\item GPT-3~\cite{GPT3} while trained on web has shown inspiring capability on synthesizing code. Models have also been explicitly trained on code with documentation~\cite{Codex, GoogleMNN, PyMT5}. In particular, Codex~\cite{Codex}, that is part of GitHub Copilot, generates Python code from natural language descriptions.  
Spider~\cite{spiderc,spiderp} is a text-to-SQL competition  where many tools compete~\cite{ratsql,spiderw}.

\item \citet{manshadi} and \citet{razaijcai} synthesize string transformations.
 {\sc WebQA}~\cite{webqa} synthesizes programs to extract information from webpages.
{\sc Regel}~\cite{regel} and~\citet{ye} synthesizes regular expressions. {\sc Mars}~\cite{mars} synthesizes data wrangling operations. These techniques have not been demonstrated at the scale of Pandas that has hundreds of operations.
\end{enumerate}

Jigsaw fixes the output of \lm\ and is hence related to work on program repair like {\sc Refazer} that learns code transformations from edits used to fix programs~\cite{Refazer}. Jigsaw's interface is inspired from B2's~\cite{B2} interface that augments visualizations to notebooks.

\section{conclusion and future work}
\label{sec:discussion}
Jigsaw is the first tool for synthesizing code for large APIs like Pandas that leverages the advancements in \lm s. The key contribution of Jigsaw lies in the post-processing steps that drastically improve the quality of the code generated by \lm s like GPT-3. 
In particular, the multimodal synthesis of Jigsaw outperforms both the baselines that exclusively use \lm s and  those that exclusively use I/O examples for program synthesis.
However, several challenges remain before we can have a true ``pair programmer'' experience with \lm s and we discuss a couple of them.

First, in this paper, the quality of the synthesized code is largely determined by the I/O examples. 
However, in practice, code quality is more nuanced than correctness on unit tests. Ideally, the synthesized code should have high performance, should not have security vulnerabilities~\cite{CodexVulnerabilitiesPaper}, and respect licensing attribution~\footnote{\url{https://www.wired.com/story/github-commercial-ai-tool-built-open-source-code/}}.

Second, Jigsaw focuses on multi-modal specifications with natural language intent and I/O examples. However, even multi-modal specifications can be weak or ambiguous, and would need to be refined using richer specifications like
preconditions, postconditions, invariants, bounds on resource usage like time and memory, etc., to obtain the intended code.

\out{
\begin{itemize}

    \item A true pair programmer would require even richer specs. Multimodal specs might not be enough.
    \item Even for correctness, we might need to go beyond test cases.
\end{itemize}
}

\out{

\begin{lstlisting}
dfout = dfin.sort_values(by =["col2","col1"], ascending =[False, True])
dfout = dfin.sort_values(by =["col2","col1"], ascending =[True, False])
\end{lstlisting}

\out{
\skanda{An on the fly AP, fixing red cases with RL/bandits?, using asserts, the actual content of dataframes, splitting/rephrasing complex queries, better UI, more libraries/languages}
}
}

\medskip
\noindent\textbf{Acknowledgement.} We thank Dhvanil Sanghvi for helping us perform a preliminary evaluation of \jigsaw\ with  Tensorflow (reported in Section 5.5), and Arjun Radhakrishna for helping us with PROSE and Refazer related queries.

\pagebreak
\bibliography{main}


\end{document}